\documentclass[11pt, aabib]{article}
\usepackage{natbib}
\usepackage{stix}

\usepackage{amsxtra, amsfonts, amssymb,latexsym}
\usepackage{amsthm,amsmath}
\usepackage{graphicx}
\usepackage[english]{babel}
\usepackage{accents}
\usepackage{hyperref}

\usepackage{pdfsync}

\newtheorem{thm}{Theorem}

\theoremstyle{definition}

\theoremstyle{definition}

\theoremstyle{remark}

\newcommand{\half}{\mbox{$\frac{1}{2}$}}
\newcommand{\beq}{\begin{equation}}
\newcommand{\eeq}{\end{equation}}

\newcommand{\x}{{\bf x}}

\newcommand{\pot}{V_{\om,k}}

\newcommand{\om}{\omega}

\newcommand{\rhoNm}{\rho_{N,m}}
\newcommand{\rhoNmone}{\rho_{N,m} ^{(1)}}

\newcommand{\HNm}{\mathcal{H}_{N,m}}

\newcommand{\MFmf}{\mathcal E ^{\rm MF}_{N,m}}

\newcommand{\rhoMFm}{\rho ^{\rm MF}_{N,m}}

\newcommand{\intR}{\int_{\R ^2}}

\newcommand{\R}{\mathbb{R}}

\begin{document}
\centerline{\bf \Large\Large Quantum Hall Phases of Cold Bose Gases
}
\bigskip\bigskip
{\large \bf Nicolas Rougerie}

{ \small Ecole Normale Supérieure de Lyon \& CNRS, UMPA (UMR 5669)\\ 
E-mail: nicolas.rougerie@ens-lyon.fr}

\bigskip\bigskip
{\large \bf Jakob Yngvason}

{\small Fakult\"{a}t f\"{u}r Physik, Universität Wien, Boltzmanngasse 5, 1090 Vienna, Austria\\
Erwin Schr\"{o}dinger Institute for Mathematics and Physics, Boltzmanngasse 9, 1090 Vienna, Austria.\\
E-mail: jakob.yngvason@univie.ac.at}

\bigskip

\bigskip\bigskip
\noindent{\bf \large Abstract}  Cold atomic gases of interacting bosons subject to rapid rotation and confined in anharmonic traps can theoretically exhibit analogues of  the fractional quantum Hall effect for electrons in strong magnetic fields. In this setting the Coriolis force due to the rotation mimics the Lorentz force on charged particles but artificial gauge fields can also be obtained by coupling the internal structure of the atoms to light fields. The chapter discusses mathematical aspects of transitions to different strongly correlated phases that appear when the parameters of a model Hamiltonian are varied.
 

\section{Key objectives}
\begin{itemize}
\item Define a model Hamiltonian for trapped, interacting  bosons in rapid rotation.
\item Describe the confinement to the lowest Landau level for the kinetic part of the Hamiltonian.
\item Discuss the yrast curve and the transition from an uncorrelated ground state in the lowest Landau level to fractional quantum Hall states.
\item Discuss transitions from a Laughlin state to other strongly correlated states including giant vortex states in anharmonic traps.
\item Derive density profiles of the strongly correlated states by means of Laughlin's plasma analogy and rigorous mean field theory.
\end{itemize}

\section{Notations and acronyms}
$\Omega_{\rm trap}$: Frequency of a harmonic trap.\\
$\Omega_{\rm rot}$: Rotational frequency.\\
$\omega=\Omega_{\rm trap}-\Omega_{\rm rot}$: Frequency difference.\\
$\mathbf L$: Angular momentum.\\
$V^{\parallel}$: Transverse confining potential.\\ 
$e^\parallel$: Spectral gap of the transverse confinement.\\
$\mathcal L=\mathbf e_3\cdot \mathbf L$: Transverse component of angular momentum.\\
LLL: Lowest Landau level.

\section{Introduction}


Since the early 1990's it has been possible to isolate ultra-cold atomic gases of bosons in magneto-optical traps and  maintain them in metastable states that can  be modelled as ground states of many-body Hamiltonians with repulsive short range interactions.
At sufficiently low temperatures the gas forms a Bose-Einstein condensate
(BEC) (\cite{Kett, Cor}) with all the atoms in the same state that can for dilute gases be described by an effective single particle equation of Gross-Pitaevskii type (\cite{PS, PiSt}).  

The atoms usually have no electric charge, but an artificial magnetic field can be imposed on them by a steady rotation of the trap. The analogy between
the Lorentz and Coriolis forces implies that the Hamiltonian for neutral
atoms in a rotating trap is formally identical to that of charged particles in a uniform magnetic field, apart from the centrifugal potential associated with the rotation.
The analogy shows up, in particular, in the emergence of quantized vortices in cold rotating
Bose gases (\cite{Bre,  A,  Cooper, Fe1, CPRY}), similar to those appearing
in type II superconductors submitted to external magnetic fields. 
  
An even more striking possibility is to employ rapid rotation to create phases with the characteristics
of the fractional quantum Hall effect (FQHE) for electrons in very large magnetic fields (\cite{PraGir-1987, ChaPie-88, ChaPie-95, StoTsuGos-99, Jain-07}).  In this regime, the Bose gas is no
longer a condensate and mean-field theories fail to capture this phenomenon. A
 truly many-body description is required and strongly correlated states, such as the celebrated
Laughlin state (\cite{Laughlin-83}), may theoretically occur. This requires, however that the Coriolis force is very strong which inevitably means that the centrifugal force is also strong and may outweigh the trapping force. This can lead to an instability of the system unless special countermeasures are taken (\cite{Bre, SCEMC, Vie, Fe1}). In fact, the competing demands of a large rotation speed on
the one hand, necessary for entering the FQHE regime, and of a sufficiently stable system
on the other hand, have so far left the FQHE regime in rotating gases unattainable with current technology. Alternative methods to create artifical gauge fields have therefore been proposed and even partly realized in experiments (\cite{DGJO, RRD, Clark, HC}). 

From a theoretical point of view, however,  the use of an anharmonic trapping potential to beat  the destabilizing effect of the Coriolis force in a rotating trap remains a conceptually simple road to strongly correlated phases of  Bosons with short range interactions.  In this chapter we shall discuss how different phases may be produced by varying the parameters in this set-up. The same conclusions will apply to other experimental realizations of the Hamiltonian whenever they can be achieved.

 \section{The basic Hamiltonian}
 
We consider first the following many-body Hamiltonian for $N$ bosons of unit mass with position vectors $\x_{(1)},\dots \x_{(N)}$ in a rotating frame of reference:
\begin{equation}
H_N= 
\sum_{i=1}^{N} \left(- \half \nabla^2_{(i)} +V(\x_{(i)})-{\mathbf L}_{(i)}\cdot {{\mathbf \Omega}}\right)+
\sum_{1\leq i < j\leq N } v(|\x_{(i)} - \x_{(j)}|).\label{ham1}
\end{equation}
 Here $\mathbf \Omega=\Omega_{\rm rot}\mathbf e_3$  is the angular velocity, with $\mathbf e_3$ the unit vector in the 3-direction and with $\Omega_{\rm rot}\geq 0$, $\mathbf L= -\mathrm i\mathbf \x\times \nabla $ the angular momentum operator, 
  \beq V(\x)=\half \Omega_{\rm trap}^2 r^2+V^{\parallel}(x_3)\eeq
  with $\x=(x_1,x_2,x_3)$, $r^2=(x_1)^2+(x_2)^2$,  a quadratic trapping potential in the 12-plane, $V^{\parallel}$ a confining positive potential in the 3-direction, and $v$ the pair interaction potential, assumed to be $\geq 0$. Units have been chosen so that Planck constant $\hbar$ as well as the particle mass is 1. 
  
It is convenient to define the gauge field as \beq \mathbf A(\x)=\Omega_{\rm trap}(x_2,-x_1,0)\eeq
and write the Hamiltonan \eqref{ham1} as
  \beq H_N=\sum_{j=1}^N\left\{\half (\mathrm i\nabla_{(j)}+\mathbf A(\x_{(j)}))^2+\omega\, \mathbf e_3\cdot \mathbf L_{(j)}+V^{\parallel}(x_3)\right\}+  \sum_{ i < j } v(|\x_{(i)} - \x_{(j)}|) \eeq
with \footnote{Alternatively, one can define $\mathbf A(\x)=\Omega_{\rm rot}(-x_2,x_1,0)$ in which case $\omega \mathbf e_3\cdot \mathbf L$ is replaced by $\half (\Omega_{\rm trap}^2-\Omega_{\rm rot}^2)r^2
$.} \beq\omega=\Omega_{\rm trap}-\Omega_{\rm rot}.\eeq

We are interested in the case that $\omega>0$ is small and $\Omega_{\rm rot}$ (hence also $\Omega_{\rm trap}$) is large.

\subsection{Confinement to the Lowest Landau Level}
\subsubsection{The 1-particle case}

The single-particle Hamiltonian
\beq H_1=\half (\mathrm i\nabla_\perp+\mathbf A(\x))^2+\omega \mathcal L-\half\partial_3^2+ V^\parallel(x_3)\label{5}\eeq
with the notation $\mathcal L=\mathbf e_3\cdot \mathbf L$ and $\nabla_\perp=(\partial_1,\partial_2)$,
 is a sum of three commuting operators,
\beq \half (\mathrm i\nabla_\perp+\mathbf A(\x))^2\label{7}\eeq
 \beq  \omega \mathcal L\label{8}\eeq
 \beq -\half\partial_3^2+ V^\parallel(x_3).\label{9}\eeq
 The first operator,  $\half (\mathrm i\nabla_\perp+\mathbf A(\x))^2$, has the form of a magnetic Hamiltonian with the {\it Landau spectrum} 
\beq (n+\half)\mathrm 2\Omega_{\rm trap},\quad n=\mathrm{0,1,2},\dots, .\eeq
The spectrum of $\omega \mathcal L$ is
\beq \ell\omega, \quad \ell=0,\pm 1,\pm 2\dots\eeq
 and $-\half\partial_3^2+ V^\parallel(x_3)$  has a spectral gap, $e^\parallel>0$, above its ground state.\smallskip

The energy of \eqref{5} is minimized in states with $n=0$, i.e., in the {\it lowest Landau level} (LLL) of the magnetic Hamiltonian. Moreover,  in the lowest energy state the motion in the $x_3$-direction is `frozen' in the ground state of $h^\parallel$. 
Confinement to the LLL in the $N$-particle context, taking the  repulsive interaction into account, is  discussed in subsection \ref{sec:N} below.

From now on we focus on the 2-dimensional motion in the 12-plane and choose units so that  $\Omega_{\rm trap}=1$.

\subsubsection{Complex notation and Bargmann space}
 It is convenient to adopt complex notation for the position variables. Replacing  $(x_1,x_2)$ by the complex coordinate $z=x_1+\mathrm i x_2$   and denoting
$\partial=\half(\partial_{1}-\mathrm i\partial_{2})$, $\bar \partial=\half(\partial_{1}+\mathrm i\partial_{2})$
we can write
\beq \half (\mathrm i\nabla_\perp+\mathbf A(\x))^2=2\left(\hat a^\dagger \hat a+\half\right)\eeq
with 
\beq \hat a^\dagger:=\half(-2\partial+\bar z),\qquad \hat a:=\half (2\bar \partial+ z).\eeq
These operators satisfy the canonical commutation relations \beq [\hat a,\hat a^\dagger]=1.\eeq
The operators 
\beq \hat b^\dagger:=\half (-2\bar \partial+ z),\qquad \hat b:=\half (2\partial+ \bar z) \eeq 
also satisfy the canonical commutation relations and commute with $\hat a$ and $\hat a^\dagger$. They correspond to a replacement $\mathbf A\to-\mathbf A$:
\beq \half (\mathrm i\nabla_\perp-\mathbf A(\x))^2=2(\hat b^\dagger \hat b+\half).\eeq
Moreover,
\beq \hat b^\dagger \hat b-\hat a^\dagger \hat a=z\partial-\bar z\bar \partial =\mathcal L.\eeq
Hence the eigenvalues of $\hat a^\dagger \hat a$ are infinitely degenerate and  the degenerate eigenstates can be labelled by  eigenvalues of  either  $\hat b^\dagger \hat b$ or  $\mathcal L$. 


 The lowest eigenvalue of $\hat a^\dagger\hat  a$ is zero so the eigenfunctions $\psi(z,\bar z)$ in the LLL are solutions of  the equation $\hat a\psi=0$, i.e.,
   \beq \bar \partial \psi(z,\bar z)=-\half z\psi(z,\bar z).\eeq
 This means that
   \beq\psi(z,\bar z)=\varphi(z)\exp(-|z|^2/2)\eeq
with $\bar \partial \varphi(z)=0$, i.e., $\varphi$ is an {\it analytic} (holomorphic) function of $z$.
   
 The LLL can thus be regarded as the {\it Bargmann space} $\mathcal B$  (\cite{Bargmann-61, GJ}) of  analytic functions $\varphi$ such that
 \beq \langle \varphi,\varphi\rangle:=\int |{\varphi(z)}|^2 \exp(-|z|^2)\,\mathrm d^2z<\infty\eeq
   where $\,\mathrm d^2z$ denotes the Lebesgue measure on $\mathbb C$ (regarded as $\mathbb R^2$).

 The Bargmann space is a Hilbert space with scalar product
    \beq \langle \varphi,\psi\rangle:=\int \bar \varphi(z) \psi(z)\exp(-|z|^2)\,\mathrm d^2z.\eeq
    On $\mathcal B$ the angular momentum operator is  $\mathcal L=z\partial$.    
    Moreover, for $\varphi\in\mathcal B$,
    \beq \langle\varphi,\mathcal L\varphi\rangle=\int ( |z|^2-1)\,|{\varphi(z)}|^2 \exp(-|z|^2)\,\mathrm d^2z .\label{19}\eeq
    The eigenvalues of $\mathcal L$ restricted to $\mathcal B$ are $\ell=0,1,2,\dots$ with corresponding normalized eigenfunctions 
   \beq\varphi_\ell(z)=\left(\pi \ell!\right)^{-1/2}\, z^\ell.\eeq
   By \eqref{19} the probability density $|\varphi_\ell(z)|^2 e^{-|z|^2}$ is localized around $|z|=\sqrt{\ell+1}$. 

 \subsubsection{The $N$-particle case, contact interaction \label{sec:N}}

For $N$ bosons in the LLL the relevant Hilbert space is 
\beq \mathcal B_N=\mathcal B^{\otimes^N_{\rm symm}},\eeq
 i.e., it consists of symmetric, analytic functions $\psi$ of $z_1,\dots,z_N$ such that
 \beq\label{144} \int_{\mathbb C^N}|\psi(z_1,\dots,z_N)|^2\exp\Big(-\sum_{j=1}^N|z_j|^2\Big)\,\mathrm d^2z_1\cdots \,\mathrm d^2z_N<\infty.\eeq 
 
  As next we take the {\it interaction}  
  $\sum_{1 \leq i < j \leq N} v(|\x_i - \x_j|)$ into account,



 For short range, nonnegative interaction potentials $v$ it was proved in (\cite{LS-09})  that  under suitable conditions on $N$, $v$ and $\omega$:
 \begin{itemize}
 \item The ground state of \eqref{5} is effectively confined to the LLL in the 12-variables and in the 3-direction to the  the ground state of \eqref{9}. 
 \item The interaction potential $v$ can be replaced by a contact interaction (pseudopotential) in the 12-variables of the form $g\delta(z_i-z_j)$ with a coupling constant
 \beq g=a\cdot \sqrt {e^\parallel}\hbox{$\int |\chi|^4$}\eeq
 where $a$ is the $s$-wave scattering length of the potential $v$ and $\chi$ the ground state wave function of \eqref{9}.
 \end{itemize}
Precise statements are contained in Theorems 1 and 2 in \cite{LS-09}. For the emergence of the pseudopotential see also (\cite{SY-20}),  Theorem 2. Since  wave functions in the Bargmann space $\mathcal B_N$ are analytic, the 2-body contact potential $\delta(z_i-z_j)$ is well defined and, in fact, given by a bounded operator:
 Defining $\delta_{12}$ on $\mathcal B_2$ by
   \beq\delta_{12}\varphi(z_1,z_2)=\frac 1{2\pi}\varphi\big(\half(z_1+z_2), \half(z_1+z_2)\big),\eeq
a simple computation, using the analyticity of $\varphi$, shows that
\beq \langle \varphi,\delta_{12}\varphi\rangle=\int_{\mathbb C}|{\varphi(z,z)}|^2\exp(-2|z|^2)\,\mathrm d^2z.\eeq
Thus the formal $\delta(z_i-z_j)$ can be replaced by the projection operator $\delta_{ij}$.
The effective Hamiltonian on the LLL can be written as
\beq\boxed{\label{yrastham} H_N^{\rm LLL}= {\omega}\,\mathcal L_N+{g}\,\mathcal I_N}\eeq 
with
\beq \mathcal L_N=\sum_{i=1}^Nz_i\partial_i\, \quad\quad \mathcal I_N=\sum_{i<j}\delta_{ij}.\eeq
We denote its ground state energy  by $E_N^{\rm LLL}(\omega,g)$.
 
 Writing $v(\x)=v_a(\x)=a^{-2}v_1(\x/a)$ and keeping $v_1$ fixed while varying $a$, we denote by
 by $E_N(\omega,a)$ the ground state energy of \eqref{ham1} with $N$ $\times$ the lowest energy of \eqref{7} and \eqref{9} subtracted. 
 Then Theorem 1 in \cite{LS-09}  states that 
\beq \lim \frac{E_N(\omega,a)}{E_N^{\rm LLL}(\omega,g)}=1
\eeq  
 in a limit where \beq (Ng\omega)^{1/2}\ll \min\{\Omega_{\rm trap}, e^\parallel\}.\eeq
 Moreover, in the same limit also the ground state converges to its projection onto the LLL (Theorem 2 in \cite{LS-09}). 

 

\subsection{The yrast curve}

An important property of the Hamiltonian \eqref{yrastham}  is that the operators $\mathcal L_N$ and $\mathcal I_N$ commute. The  lower boundary of (the convex hull of) their joint spectrum in a plot with angular momentum as the horizontal axis is called the  {\it yrast curve}. See Fig. 4 and  \cite{Vie} for its qualitative features. 

As a function of the eigenvalues $L$ of $\mathcal L_N$ the yrast curve $I(L)$ is decreasing from $I(0)= (4\pi)^{-1} N(N-1)$ to $I(N(N-1))=0$. The monotonicity follows from the observation that if a simultaneous eigenfunction of $\mathcal L_N$ and $\mathcal I_N$ is multiplied by the center of mass, $(z_1+\cdots+z_N)/N$, the interaction is unchanged while the angular momentum increases by one unit. 

For a given ratio $\omega/g$ the ground state of \eqref{yrastham} (in general not unique) is determined by the point(s) on the yrast curve  where a supporting line of the curve has slope $-\omega/g$. The ground state energy is
\beq E(N,\omega,g)=\min_L(\omega L+g\,I(L)).\eeq

The {\em filling factor} of a state with angular momentum $L$ is defined as
\beq \nu=\frac{N(N-1)}{2L}=\frac N{N_{\rm v}}\eeq
where $N_{\rm v}=2L/(N-1)$ is called the number of vortices see  \cite{Cooper}, Sec. 2.4.
The filling factor of the ground state depends on the ratio $\omega/g$ and varies from $\infty$ to 0 as the ratio decreases and the angular momentum increases. Note the difference to the fermionic case, where the maximal filling factor in a fixed Landau level is 1.

\subsubsection{Spectral gaps \label{spectralgap}}

For every value of the angular momentum $L$ the interaction operator $\mathcal I_N$ has a nonzero spectral gap above 0
\beq \Delta_N(L)=\inf \{{\rm spec}\ \mathcal I_N\upharpoonright_{\mathcal L_N=L}\setminus \{0\}\}>0.\eeq
This follows simply from the fact that states with a given eigenvalue $L$ correspond to the finite dimensional space of symmetric homogeneous polynomials of degree $L$ and contains states with strictly positive interaction energy. (Take for example $\hbox{(const.)}\sum_j z_j^L$.)
The gap, and hence the Yrast curve $I(L)$, are  monotonously decreasing with $L$ for the reason already mentioned: The angular momentum of an eigenstate of $\mathcal I_N$ can be increased by one unit by multiplying the wave function with the center of mass coordinate $(z_1+\cdots +z_N)/N$. 
This does not change the interaction energy and leads to a family of `daughter states' for each state on the yrast curve.
There is numerical (see e.g. \cite{VHR, RCJJ}) and some theoretical evidence that
\beq\label{157} \Delta_N(L)\geq \Delta_N(N(N-1)-N)=\min_{L'\leq N(N-1)}\Delta_N(L')=:\Delta >0\eeq
for all $L$ with $\Delta$ {\it independent of $N$} but this is still not proved. (Proofs  in a simplified setting  can be found in \cite{NacWarYou-20b, WYa, WYb}.) We shall call the validity of \eqref{157} the {\it spectral gap conjecture}.

\begin{figure}[ht]
\center
\fbox{\includegraphics[width=12cm, bb=100 450 500 730]{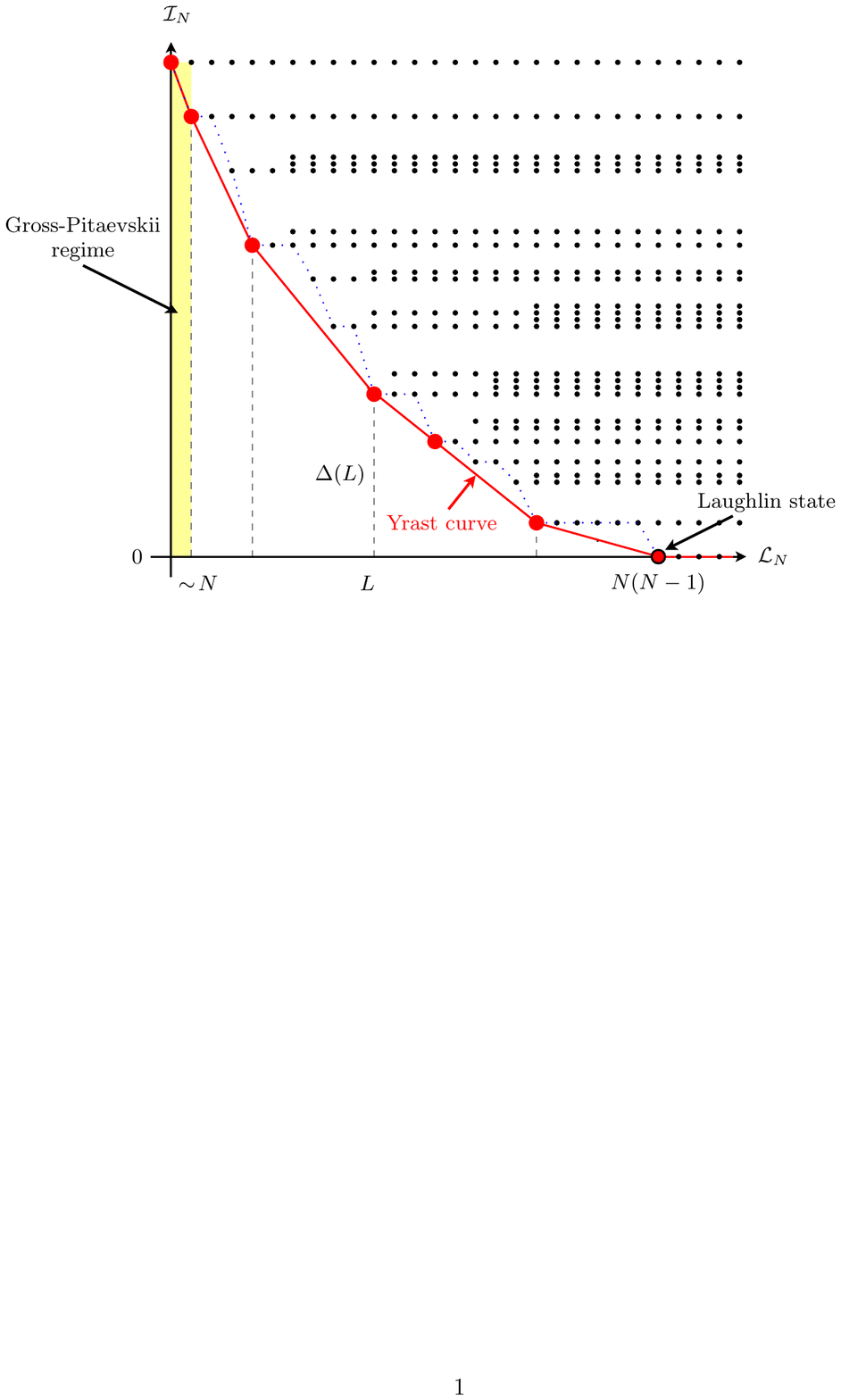}}
\caption{The joint spectrum of $\mathcal L_N$ and $\mathcal I_N$ (adapted from \cite{LS-09}).}
\label{fig4}
\end{figure}



\subsubsection{The uncorrelated (Gross-Piaevskii) regime}

A rough estimate for the radius $R$ of the system, assuming that the kinetic and interaction energy of \eqref{yrastham} are of the same order of magnitude, gives \beq R\sim (Ng/\omega)^{1/4}\qquad \hbox{\rm and}\qquad 
L\sim N\Omega R^2\sim N(Ng/\omega)^{1/2}.\eeq
 Thus, if If $\omega/g \gg N^{-1}$ then $L\ll N^2$ and $\nu\gg 1$. 
In this case the ground state can be shown to be well described by an uncorrelated product state
  $\left(\varphi^{\rm GP}\right)^{\otimes N}$ where $N^{1/2}\varphi^{\rm GP}$ minimizes the {\it Gross-Piitaevskii energy functional on Bargmann space} (\cite{ABN2, AB})
\beq\mathcal E^{\rm GP}[\varphi]=\omega\langle\varphi, \mathcal L\varphi\rangle+\frac g2\int_{\mathbb C}|\varphi(z)|^4\exp(-2|z|^2)\,\mathrm d^2z\eeq
under the condition $\int|\varphi|^2\exp(-|z|^2)\mathrm d^2z=N$ with energy \beq E^{\rm GP}_N(\omega, g)=NE_1^{\rm GP}(\omega, Ng).\eeq

More precisely, the following holds (\cite{LSY}):
 
\begin{thm}[GP limit theorem in LLL]  
    For every  $c>0$ there is a $C<\infty$ such that 
  \beq E^{\rm GP}_N(\omega,g)\geq E_N(\omega,g)\geq E_N^{\rm GP}(\omega, g)(1-C(g/N\omega)^{1/10})\eeq
  provided $gN/\omega>c$.\label{GPlim}\end{thm}
  
    The lower bound covers the whole regime $L\ll N^2$, i.e., $\nu\gg 1$, but the GP description might have a wider range of applicability, see \cite{Fe1} Section  V C.

  The proof of \eqref{GPlim} uses similar techniques as in \cite{LSe} for the 3D GP limit theorem at fixed $\Omega$ and $Na$, in particular coherent states.
  
The analysis of the GP problem defined above is of interest in its own right~\cite{AB2,ABN2,GGT}. It is expected (see~\cite{A,Vie,Cooper,Fe1} and references therein) that for $\omega \to 0$ a vortex lattice is nucleated in the condensate. Namely, the zeroes of the analytic part of the GP minimizer in Bargmann space arrange on a triangular lattice in the bulk of the condensate, with a deformation at the edge~\cite{ABD,Ho,AB2,ABN2}. Taking this fact into consideration in a local density approximation, one expects that
\beq
E_1^{\rm GP}(\omega, Ng) \sim E^{\rm TF} (\omega,Ng) \mbox{ in the limit } \omega \to 0 \mbox{ with } \omega \ll Ng \ll \omega^{-1}
\eeq
where 
\beq E^{\rm TF} (\omega,Ng) = \min\left\{ \int_{\R^2}\left( \left(\omega |x| ^2 - 1 \right) \rho (x) + \frac{Ng}{2} e^{\rm Ab} \rho(x)^2\right)dx \, | \,  \rho \geq 0, \int \rho = 1\right\} \eeq
with 
\beq
e^{\rm Ab} \approx 1.16
\eeq
the Abriskosov constant~\cite{Abr,ABN2}, i.e. the energy per area of a homogeneous vortex lattice. One also expects for the correspoing density of the GP minimizer
\begin{equation}\label{eq:dens asymp}
|u^{\rm GP}| ^2 \simeq \rho^{\rm TF} = \frac{1}{Ng e^{\rm Ab}}\left(\lambda - \omega|x|^2 \right)_+ 
\end{equation}
in the same regime, where $\lambda$ is a chemical potential ensuring normalization. The above expectations were rigorously proven in~\cite{NR} (under the restricted assumption that $Ng \ll \omega^{-3/5}$). 

These results suggest a formula distribution of quantized vortices in the condensate. Indeed\footnote{Putting the following discussion on the same level of mathematical rigor as~\eqref{eq:dens asymp} remains a challenging open problem.}, writing the GP minimizer in the lowest Landau level in the manner
\beq u^{\rm GP} (z)= c \prod_{j=1} ^m (z-a_j) e^{-\frac{|z|^2}{2}}\eeq
one finds~\cite{ABD,Ho} 
\beq \sum_{j= 1} ^m \delta_{a_j} = \frac{1}{4\pi} \left( 4 + \Delta \log |u ^{\rm GP}| ^2\right)\eeq
for the empirical distribution of the zeroes of $u ^{\rm GP}$, i.e., the locations of quantized vortices. Inserting the approximation~\eqref{eq:dens asymp} one finds 
\beq 
\sum_{j= 1} ^m \delta_{a_j} \simeq \frac{1}{\pi} + \frac{1}{4\pi} \Delta \log \left((\lambda - \omega |x|^2)_+\right).
\eeq
The first term, corresponding to a uniform density of vortices, dominates the asymptotics. The second one is a inhomogeneous correction, manifest mostly at the edge of the atomic cloud, see~\cite{ABD,AB2,ABN2} for more details.

The breakdown of the GP approximation is indicated by the number of vortices nucleated in the gas exceeding the total particle number. The above considerations indicate that this takes place when $g\gtrsim N\omega$, which marks the transition to a series of correlated many-body states, ending with a bosonic Laughlin function, as we discuss next.

  \subsection{Passage to the Laughlin state}

As the filling factor decreases the ground state becomes increasingly correlated. The exact ground states are largely unknown (except for $L\leq N$ (\cite{PB})), but candidates of states with various rational filling factors ({\it composite fermion states, Moore-Read states, Read-Rezayi states,...}) for energy upper bounds have been suggested and studied as discussed in the review article \cite{Cooper}. 

If $\omega/g<\Delta/N^2$ one reaches the {\em Laughlin state} with filling factor $\half$ whose wave function in Bargmann space is
\beq \psi_{\rm Laughlin}(z_1,\dots,z_N)=c\prod_{i<j}(z_i-z_j)^2.\eeq
It has interaction energy 0 and angular momentum $L=N(N-1)$.  \smallskip
For an intuitive picture of the Laughlin state the following analogy may be helpful: The density $|\Psi_{\rm Laugh}(z_1,...z_N)|^2$ assigns probabilities to the possible configurations of $N$ points moving in the plane. The points like to keep a distance at least of order 1 from each other because the factors $|z_i-z_j|^4$ strongly reduce the probability when the particles are close. On the other hand the damping due to the gaussian favours a tight packing of the `balls' of size $O(1)$ around the individual particles.  The motion is strongly correlated in the sense that if one ball moves, all the other have also to move in order to satisfy these constraints.


\section{Adding an anharmonic potential}

The limit $\omega\to 0$, keeping $\omega>0$, is experimentally very delicate.
 For  stability,  but also to study new effects, we consider now a modification of the Hamiltonian \eqref{yrastham} by adding a small anharmonic term:
\beq H^{\rm LLL}_N\rightarrow H^{\rm LLL}_{k,N}:=H^{\rm LLL}_N+k\sum_{i=1}^N |z_i|^4\eeq
with a new parameter $k>0$.  The potential $|z|^4$  can be expressed through $\mathcal L$ and $\mathcal L^2$ on Bargmann space, because
with $\mathcal L=z\partial$ we have by partial integration, using the analyticity of $\varphi$:
\beq \langle\varphi,\mathcal L\varphi\rangle=\int |{\varphi(z)}|^2 (|z|^2-1)\exp(-|z|^2)\,\mathrm d^2z \eeq 
 and
\beq \langle\varphi,\mathcal L^2\varphi\rangle=\int (|z|^4-3|z|^2+1)|{\varphi(z)}|^2 
 \exp(-|z|^2)\,\mathrm d^2z. \eeq 
Thus the modified Hamiltonian, denoted again by $H^{\rm 2D}_N$,  can be written (up to an unimportant additive constant)
\beq \boxed{\label{162}H^{\rm LLL}_{k,N}=(\omega+{3k})\mathcal L_N+{k} \sum_{i=1}^N \mathcal L_{(i)}^2+g\,\mathcal I_N.}\eeq
 
\subsection{Fully correlated states}

The Bargmann space $\mathcal B^N$ with the scalar product \eqref{144} is naturally isomorphic to the Hilbert space 
$\mathcal H^N_{\rm LLL} \subset L^2(\mathbb C)^{\otimes_s N}, \mathrm d^{2N}z)$ 
consisting  of wave functions of the form 
\beq \Psi(z_1,\dots,z_N)=\psi(z_1,\dots,z_N)\exp(-\sum_j|z_j|^2/2), \qquad \psi\in\mathcal B_N\eeq with the standard $L^2$ scalar product. The energy can accordingly be  considered as a functional on this space, 
\beq \mathcal E[\Psi]=\int V_{\omega,k}(z)\rho_\Psi^{(1)}(z)+\langle\Psi,\mathcal I_N\Psi\rangle,\eeq
where $\rho_\Psi^{(1)}$ is the  one-particle density of $\Psi\in \mathcal H^N_{\rm LLL}$ with the  normalization 
$\int \rho_\Psi^{(1)}(z)\,\mathrm d^2z=N$
and the potential is
\begin{equation}
\pot (z) = \om\, |z|^2 + k\,|z|^4. 
\end{equation}
Note that now $\omega<0$ is allowed, provided $k>0$. 

We shall call states with vanishing interaction energy, i.e., $\Psi\in \mathrm{ker}\, \mathcal I_N$ {\em fully correlated}, because the particles stay away from each other in the sense that the wave function vanishes if $z_i=z_j$ for some pair $i\neq j$, in sharp contrast to a fully {\it un}correlated Hartree state. The fully correlated states  in $\mathcal H^N_{\rm LLL}$ are of the form
\beq \Psi(z_1,...z_N)=\phi(z_1,\dots,z_N)\Psi_{\rm Laugh}(z_1,...z_N)\eeq 
with $\phi$ symmetric and analytic, and the Laughlin state
\beq \Psi_{\rm Laugh}(z_1,...z_N)= c \prod_{i<j} \left( z_i -z_j\right) ^2 e ^{-\sum_{j=1} ^N |z_j| ^2 / 2 }.\eeq


For the Hamiltonian {\it without} the anharmonic addition to the potential the Laughlin state is an {\it exact} fully correlated ground state with energy 0 and 
angular momentum $L_{\rm Laugh}=N(N-1)$. 
This is {\it not} true for $k\neq 0$ because
$ \sum_{i=1}^N \mathcal L_{(i)}^2$ does not commute with $\mathcal I_N$.
Note, however, that $\mathcal L_N$ still commutes with the Hamiltonian.\smallskip

We now address the following question: Under what conditions is it possible to tune the parameters $\omega$, $g$ and $k$  so that a ground state $\Psi_0$ of \eqref{162} becomes fully correlated for $N\to\infty$? The following theorem, proved in \cite{RSY} (see also \cite{RouSerYng-13b}), gives sufficient conditions for this to happen. In order to state it as simply as possible we shall assume the `spectral gap conjecture' of Subsection \ref{spectralgap}. This conjecture is not really needed, however, because is possible to replace the assumed universal gap $\Delta$ by other gaps depending on $N$, $\omega$ and $k$, cf. Eq. (IV.5) in \cite{RSY}.
\begin{thm}[Criteria for full correlation] 
 With 
$P_{({\rm Ker\,}\mathcal I_N)^\perp}$ 
the projector onto the orthogonal complement of 
${\rm Ker\,}\mathcal I_N$
 we have 
 \begin{equation}
\left\Vert P_{({\rm Ker\,}\mathcal I_N)^\perp}\Psi_0 \right\Vert  \to 0  
\end{equation}
in the limit $N\to \infty$, $\om,k\to 0$ if one of the following conditions holds:
\begin{itemize}
\item $\om \geq 0$ and
$\om N ^2 + k N ^3 \ll g \: \Delta$.\vskip.5cm
\item $0 \geq \om \geq - 2 k N$ and
$N (\om ^2/{k}) + \om N ^2 + k N ^3 \ll g \: \Delta
$.\vskip .5cm
\item $\om \leq - 2 k N$, $|\omega|/k \lesssim N^2$ and
$ k N ^{3} \ll g \: \Delta$\vskip .5cm
\item $\om \leq - 2 k N$,  $|\omega|/k \gg N^2$ and 
$ |\om| N \ll g \: \Delta$\vskip.2cm
\end{itemize}
\end{thm}

Note: For $k=0$ the first item is just the sufficient condition for the passage to the Laughlin state, $\omega/g<\Delta/N^2$, 
while the other conditions are void because $\omega<0$ is only allowed if $k>0$.


A proof of the Theorem is given in Section V of \cite{RSY}. It is based on two estimates:
\begin{itemize}
\item A lower bound for the ground state energy at fixed angular momentum $L$:
\beq  E_0(L)\geq (\omega+3k)L+k\frac {L^2} N.\label{lb}\eeq
\item An upper bound for the energy of suitable trial functions.
\end{itemize}
\medskip

The first bound \eqref{lb} is quite simple; it follows essentially from the inequality
\beq \sum_i \mathcal L_{(i)}^2\geq \frac 1N\left (\sum_i \mathcal L_{(i)}\right)^2\eeq
 that holds because $\mathcal L_{(i)}$ and $\mathcal L_{(j)}$ commute for any $i,j$.


The upper bound is achieved by means of  trial states of the form `giant vortex times Laughlin', namely,
with $m\geq 0$ and $c_{m,N}$ a normalization constant,
\beq  \Psi_{\rm gv}^{(m)}(z_1,\dots,z_N) = c_{m,N} \prod_{j=1} ^N z_j ^m \prod_{i<j} \left( z_i - z_j\right) ^2 e^{-\sum_{j=1} ^N |z_j| ^2 / 2}\label{ub} \eeq 
These are special `quasi hole' states (\cite{Laughlin-83})  but for $m\gtrsim N$, i.e., $mN\gtrsim N^2$, the label `giant vortex' appears more appropriate. 
\footnote{Note that mathematically and physically these states are quite different from the  {\it uncorrelated} giant vortex states in the GP theory of rotating Bose gases discussed in \cite{CPRY}.}



The energy of a trial state of the form \eqref{ub}  can be estimated using properties of the angular momentum operators and the radial symmetry in each variable of the gaussian measure with the pre-factor $\prod_{j=1} ^N |z_j| ^{2m}$.  The computation is presented in Eqs. (V3)-(V7) in \cite{RSY}. Optimizing the estimate over $m$ leads to

\begin{equation}\label{eq:intro m opt}
m_{\rm opt} = \begin{cases}
               0& \mbox{\rm if } \om \geq - 2 k N \\
               \frac{|\om|}{2 k}- N &\mbox{\rm if } \om < - 2 k N.
              \end{cases} 
\end{equation}
\medskip

This is consistent with the picture that the Laughlin state is an approximate ground state in the first two cases of Theorem 1, in particular for negative $\omega$ as long as  $|\omega|/k\lesssim N$. The angular momentum remains $O(N^2)$ in these cases.



When $\omega<0$ and  $|\omega|/(kN)$ becomes large  the angular momentum is approximately  \beq L_{\rm qh}:=N|\omega|/k\gg N^2,\eeq much larger than for the Laughlin state.
A further transition  at $|\omega|/k\sim N^2$ is manifest through the change of the subleading contribution to the energy of the trial functions. Its order of magnitude changes from $O(kN ^3)$ to $O(|\om| N)$ at the transition.\smallskip

To gain further insights into the physics of the transition we consider the particle density of the trial wave functions. This analysis (\cite{RouSerYng-13b}) is based on Laughlin's \lq plasma analogy\rq\  where the $N$-particle density is written as a Gibbs distribution of a 2D Coulomb gas \cite{Laughlin-83}. This distribution can be studied in a rigorous mean field limit which brings out the essential features of the single particle density for large $N$.



\subsection{The $N$-particle Density as a Gibbs Measure}

We denote $(z_1,\dots,z_N)$ by $Z$ for short and consider the scaled $N$ particle density (normalized to 1)
\beq \rhoNm (Z) := N ^N \left| \Psi_{\rm gv}^{(m)} (\sqrt{N} Z )\right| ^2.\eeq
We can write
\begin{eqnarray*}\hskip-.1cm
\rhoNm (Z) &=& \mathcal Z_{N,m}^{-1} \exp\left( \sum_{j=1} ^N \left( - N  |z_j| ^2 + 2 m \log |z_j|\right)  + 4 \sum_{i<j} \log |z_i - z_j|\right) 
\\  &=& \mathcal Z_{N,m}^{-1} \exp\left( -\frac{1}{T} \HNm(Z) \right),
\end{eqnarray*}
with $T=N^{-1}$ and 
\beq \HNm(Z)=\sum_{j=1} ^N \left(   |z_j| ^2 - \frac{2 m}N \log |z_j|\right)  - \frac 4N \sum_{i<j} \log |z_i - z_j|.\eeq


\subsubsection {Plasma analogy and mean field limit}

The Hamilton function $\HNm(Z)$ is that of a classical 2D Coulomb gas (`plasma', one-component `jellium') in a uniform  background of opposite charge and with a point charge $(2m/N)$ at the origin, corresponding respectively to the $|z_i|^2$ and the $-\frac{2 m}N \log |z_j|$ terms. \medskip

The probability measure $\rhoNm (Z)$ minimizes the free energy functional
\beq \mathcal F(\rho)=\int_{\mathbb R^{2N}} \HNm(Z) \rho(Z)+T\int_{\mathbb R^{2N}}\rho(Z)\log\rho(Z)\eeq
for this Hamiltonian at $T=N^{-1}$. \medskip

The $N\to\infty$ limit is in this interpretation a {mean field limit} where at the same time $T\to 0$. It is thus not unreasonable to expect that for large $N$, and in an appropriate sense,
\beq \rhoNm\approx \rho^{\otimes N}\eeq
with a {one-particle} density $\rho$ minimizing a {mean field free energy functional.}



The {mean field free energy functional} is defined as
\begin{equation}
\MFmf [\rho]: = \intR  W_m\,  \rho - 2 \intR\intR \rho(z)\log|z-z'|\rho(z') + N ^{-1} \int_{\R ^2} \rho \log \rho 
\end{equation}
with
\begin{equation}
W_m (z) = |z| ^2 - 2 \frac{m}{N} \log |z|.
\end{equation}
It has a minimizer $\rhoMFm$ among probability measures on $\mathbb R^2$ and this minimizer is in (\cite{RouSerYng-13b})  proved to be a good approximation for the scaled 1-particle density of the trial wave function, i.e., for
$$\rhoNmone (z):= \int_{\R ^{2(N-1)}} \rhoNm (z,z_2,\ldots,z_N) \mathrm d^2z_2\ldots \mathrm d^2z_N.$$
(Recall the scaling: This density in the scaled variables $z$ is normalizes so that its integral is 1. The corresponding density in the physical, unscaled variables $\zeta=\sqrt N z$ has total mass $N$.)

\subsubsection{Formulas for the mean field density}

The picture of the 1-particle density that arises from asymptotic formulas for the mean-field density, valid for large $N$,  is as follows:\medskip

If $m\leq N^2$, then  $\rho^{\rm MF}_m$ is well approximated by a density $\hat \rho^{\rm MF}_m$ that minimizes the mean field functional without the entropy term $N ^{-1} \int_{\R ^2} \rho \log \rho $.
This density takes a {\em constant value}  $(2\pi)^{-1}$  on an annulus with inner and outer radii (in the scaled variables!)
$$R_-=(m/N)^{1/2},\qquad R_+=(2+m/N)^{1/2}$$ and is zero otherwise.  The constant value is a manifestation of the {\it incompressibility of the density} of the trial state. \footnote{Note that these is a statement about the mean field density that is a good approximation, suitably averaged, to the true 1-particle density for large $N$. See \cite{Ciftja} for numerical calculations of the true density of the Laughlin state for $N=400$.}
\medskip

For $m\gtrsim  N^2$ the entropy term becomes important and may dominate the electrostatic interaction term $\int\int\rho(z)\log|z-z'|\rho(z')$. The density is well approximated by a gaussian profile  $\rho^{\rm th}(z)\sim |z|^{2m}\exp(-N|z|^2)$ that is centered around $(m/N)^{-1/2}$ but has maximal value $\sim N/m^{1/2}\ll 1$ for $m\gg N^2$.



Thus, as the parameters $\omega$ and $k$ tend to zero and $N$ is large, the qualitative properties of the optimal trial wave functions exhibit different phases:\medskip
\begin{itemize}
\item For positive $\omega$, more generally for $\omega/k > -2N$, the Laughlin state is the ground state.

\item When $\omega$ is negative and $|\omega|$ exceeds $2kN$ the  state changes from a pure Laughlin state to a modified Laughlin state with a `hole' in the density around the center.\medskip

\item A further transition is indicated at $|\omega|\sim kN^2$. The density profile changes from being `flat' to a Gaussian.
\end{itemize}


\begin{figure}[ht]
\center
\fbox{\includegraphics[width=12cm, bb=100 450 500 780]{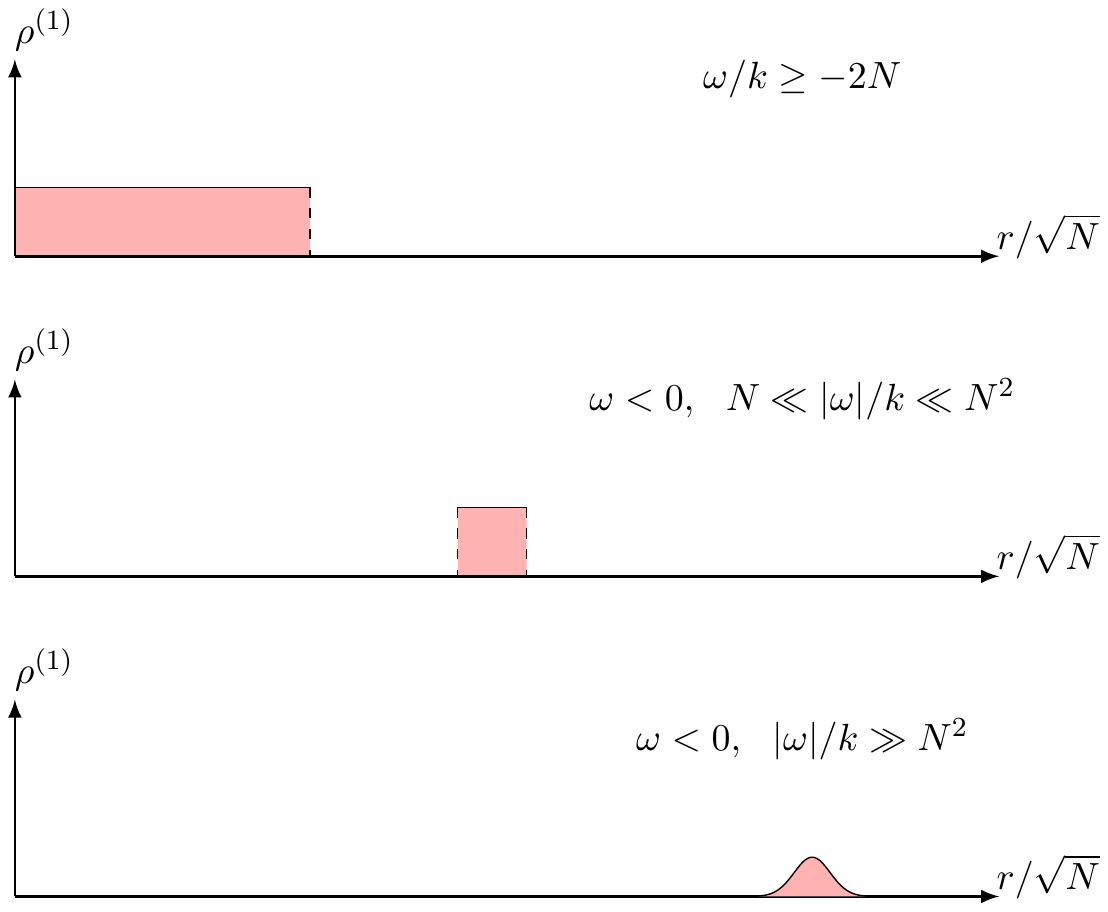}}
\caption{The three phases of the density $\rho^{(1)}_{N,m}$ (not to scale).}
\label{fig5}
\end{figure}

An intuitive understanding of these transitions may be obtained by employing the previous picture of the points $z_i$ as being the centers of essentially non overlapping balls of size $O(1)$. From the point of view of the plasma analogy the particles stay away from each other because of the repulsive Coulomb potential between them, while the attractive external potential due to the uniformly charged background keeps them as close together as possible. Modifying the wave function by a factor $\prod_j z_j^m$ has the effect of a repulsive charge of magnitude $m$ at the origin that pushes the particles (collectively!) away from the origin, creating a `hole'. The effect of such a hole on the energy of the wave function in the trap potential is to increase the energy if $\omega$ is positive. Hence the ground state will not have a hole. If $\omega<0$ the effective trapping potential has a Mexican hat shape with a minimum away from the origin, but since no ball can move without `pushing' all the other balls, it is too costly for the system to take advantage of this as long as $k$ stays above the critical value $|\omega|/2N$. For smaller $k$ a hole is formed. The balls remain densely packed until the minimum of the Mexican hat potential moves so far from the origin that an annulus of width $O(1)$ at the radius of the minimum can accommodate all $N$ balls. This happens for $k\lesssim  |\omega|/N^2$. For smaller $k$ (larger radius) the balls need not be tightly packed in the annulus and the average local density decreases accordingly.

\subsection{Summary and Conclusions}

The main conclusion from the analysis presented above of the ground states in the lowest Landau level generated by rapid rotation can be summarized as follows:

\begin{itemize}
\item For the Hamiltonan \eqref{yrastham} the parameter regime $g\ll N\omega$, i.e., $L\ll N^2$, can be described by Gross-Piaevskii theory in the LLL.
\item To enter the `fully correlated' regime with $L\geq N(N-1)$ we have studied the Hamiltonian \eqref{162} with a quadratic plus quartic trap  where the rotational frequency can exceed the frequency of the quadratic part of the trap, i.e, the frequency difference $\omega$ can be  negative. 

\item Through the analysis of  trial states for energy upper bounds, and simple lower bounds, we have obtained criteria for the ground state to be fully correlated in an asymptotic limit. The lower bounds, although not sharp, are of the same order of magnitude as the upper bounds.

\item The particle density of the trial wave functions can be analyzed using Laughlin's plasma analogy combined with rigorous mean field limits. The character of the ground state density changes at $|\omega|/k=O(N)$ and again at $|\omega|/k=O(N^2)$.

\end{itemize}


\bigskip

\noindent\textbf{Acknowledgments.} 
Funding from the European Research Council (ERC) under the European Union's Horizon 2020 Research and Innovation Programme (Grant agreement CORFRONMAT No 758620) is gratefully acknowledged.


\end{document}